# Monte Carlo Simulation of Single-Crystalline PbSe Nanowire Thermal Conductivity Using First-Principle Phonon Properties

Lei Ma [1,2], Riguo Mei[1,2], Mengmeng Liu[3], Xuxin Zhao[2], Qixing Wu[2], Hongyuan Sun[2,*]

[1] Key Laboratory of Optoelectronic Devices and Systems of Ministry of Education and Guangdong Province, Shenzhen University, 518060, People's Republic of China

[2] Key Laboratory of New Lithium-ion Batteries and Mesoporous Materials of Shenzhen City, College of Chemistry and Environmental Engineering, Shenzhen University, 518060, People's Republic of China

[3] Department of Radiology, University of California, 94107, United States

*Corresponding author, Email: hysun163@163.com

ABSTRACT:

Prior experimental studies showed that nanowires are promising structures for improving the thermoelectric performance of practical thermoelectric materials due to the strongly induced phonon-boundary scattering. However, few studies examined the impact of phonon-boundary scattering on the thermal conductivity of thermoelectric nanowires from a first-principle approach. In this work, we systematically study the role of phonon-boundary scattering with different boundary specularities on the thermal conductivity of PbSe nanowires by rigorously solving the full phonon Boltzmann transport equation without any adjustable parameters. We observe significant thermal conductivity reduction for rough PbSe nanowires with diameters less than a few hundred nanometers. The reduction reaches ~ 40% for 10 nm thick rough PbSe nanowires at room temperature. The diameter-dependent thermal conductivities are found to contain important information about the phonon mean free path distribution from a standard reconstruction algorithm. The simulation results are important for fundamental understanding of nanoscale thermal transport in thermoelectric materials and will guide future design of thermoelectric devices to achieve better energy conversion efficiency.

1. Introduction

Solid-state thermoelectric devices that are able to convert heat directly to electricity without any intermediate processes or moving parts are promising technologies for reliable and pollution-free power generation and waste heat recovery in many industrial applications[1-4]. The heat-to-electricity conversion efficiency is determined by a dimensionless thermoelectric figure of merit $ZT$ that is a function of three material parameters: $ZT = \frac{S^2 \sigma}{k} T$, where $S$ is the Seebeck coefficient, $\sigma$ is the electrical conductivity, $k$ is the total thermal conductivity that includes the contributions from both the lattice and the charge carriers, and $T$ is the average device operating temperature. In general, it is challenging to improve $ZT$ due to the fact that material's thermal and electrical properties are normally interdependent in such a way that optimizing one property typically adversely affects other properties[5]. Fortunately, one effective strategy, nanostructuring, was proposed by Hicks and Dresselhaus in 1993 to decouple the thermal conductivity from the electrical properties[6,7]. Thermoelectric nanomaterials typically have dense interfaces or material boundaries that can selectively scatter lattice vibrations while not negatively impacting the electrical properties[8]. Increased phonon scattering at interfaces and boundaries can significantly reduce the lattice thermal conductivity by shortening the effective phonon mean free path (MFP).

The effectiveness of nanostructuring was confirmed by numerous theoretical and experimental investigations on many low dimensional material systems, including quantum dots, thin films, superlattices, nanowires, to name a few[9-15]. In particular, nanowires have been of renewed research interest due to its unique properties as candidate thermoelectric materials[16–20]. For example, the measured thermal conductivities of vapor-liquid-solid-grown silicon nanowires with diameter in the a few tens to ~100 nm were much lower than the bulk thermal conductivity across a wide range of temperatures[16]. A subsequent work by Boukai *et al*.[19] observed a peak ZT of ~ 1.0 at 200K for silicon nanowires of ~ 20 nm in diameter. Hochbaum *et al*.[18] reported high performance, scalable rough silicon nanowire arrays as efficient thermoelectric materials and found that the peak ZT reaches ~ 0.6 for rough silicon nanowires of ~ 50 nm in diameter, much higher than the performance based on bulk silicon. The performance enhancement of silicon nanowires for thermoelectric applications mainly results from the substantially enhanced phonon scattering with nanowire boundaries[1]. In addition to silicon nanowires, a few other thermoelectric nanowires were also studied to examine their thermoelectric performance[21–23]. For example, Liang *et al*.[21] experimentally observed significant thermal conductivity reduction in the solution-phase synthesized p-type PbSe nanowires 50 ~ 100 nm in diameter across a wide range of temperatures. Roh *et al*.[22] found that the thermal conductivity of single-crystalline PbTe nanowire of ~ 182 nm in diameter was reduced by approximately 50% compared to its bulk value at room temperature. In a subsequent work, Li *et al*.[23] used first-principle calculations to determine the diameter-dependent thermal conductivities of $Mg_2Si_xSn_{1-x}$ alloys and observed ~ 50% reduction in thermal conductivity for nanowires with diameters small than 20 nm.

Despite of the aforementioned calculation work on $Mg_2Si_xSn_{1-x}$ nanowire system, few studies have focused on investigating the size-dependent thermal conductivities of other important thermoelectric nanowire systems. In particular, bulk PbSe is a good thermoelectric material in the mid- to high-temperature range applications[3]. In this work, we examine the potential of single-crystalline PbSe nanowires to further improve the thermoelectric energy conversion efficiency through size-effect engineering. We systematically study the role played by phonon-boundary scattering and boundary specularity on the thermal conductivity of PbSe nanowires by rigorously solving the multidimensional phonon BTE using a recently developed variance reduced Monte Carlo (MC) algorithm[24,25]. We incorporate the full phonon dispersion relation and spectral lifetimes obtained from first-principles density function theory calculations as input in the MC simulations without any adjustable parameters[26]. Both square and circular nanowires are examined and we find very weak dependence of the PbSe nanowire thermal conductivity on the nanowire shape. Approximately 40% reduction in thermal conductivity is observed for rough PbSe nanowire of 10 nm in diameter. The boundary specularity is found to increase the nanowire thermal conductivity, consistent with our understanding of specular phonon reflection. The diameter-dependent thermal conductivities are also used to reconstruct the heat-carrying phonon MFP distribution that contributes to the PbSe thermal conductivity using a recently proposed convex optimization algorithm. We observe that the size-dependent nanowire thermal conductivities contain important information about the distribution of phonon MFPs and this provides an alternative venue to experimentally measure the phonon MFP spectrum. Our simulation results will provide useful guidance for further optimizing the thermoelectric energy conversion efficiency using low dimensional materials.

2. Simulation Details

To calculate the nanowire thermal conductivity from an ab-initio approach, we must solve the full phonon BTE for the geometry of interest[5]. The phonon BTE in its full spectral form is given by:

$$\frac{\partial e_\omega}{\partial t} + \vec{V_\omega} \cdot \nabla e_\omega = -\frac{e_\omega - e_{\omega 0}}{\tau_\omega} \quad (1)$$

where $e_\omega = \hbar\omega(f_\omega - f_\omega^{eq})$ is the deviational phonon energy distribution, $e_{\omega 0} = \hbar\omega(f_{\omega 0} - f_\omega^{eq})$ is the equivalent equilibrium deviational energy distribution, $\hbar\omega$ is the phonon energy, $f_\omega$ is the phonon distribution function, $f_\omega^{eq}$ is the Bose-Einstein distribution at the specified reference temperature $T_{eq}$, $V_\omega$ is the phonon group velocity, and $\tau_\omega$ is the spectral phonon lifetime. Note that Eq. (1) omits the subscript for phonon polarization, but it applies to all the phonon branches, including acoustic and optical branches. Equation (1), given in the deviational form relative to a reference equilibrium distribution[24,25], models the evolution of the local phonon energy distribution as a function of time and space and applies to both diffusive and non-diffusive transport regimes.

In general, it is challenging to solve the multidimensional integro-differential BTE since the phonon distribution function is a scalar in the six-dimensional phase space. In particular, analytical or semi-analytical solutions only exist for simple transport geometries[27–31]. Various numerical techniques, such as finite difference and finite volume approaches, have also been developed[32–34]. However, applying those techniques involve significant computational cost, especially for complicated simulation geometries incorporating the full spectral phonon properties as input. Fortunately, recent improvement in computational efficiency through the linearized deviational energy-based BTE in the linear response regime enables the simulations of complicated geometries using the stochastic Monte Carlo approach with high accuracy[24,25,35,36]. The key idea of the linearized deviational energy-based BTE is two-fold. On one hand, by observing the fact that the temperature difference in typical nanoscale thermal transport experiments is very small, we can simulate only the deviation from a chosen reference state[24]. The solution of the reference state is precisely known beforehand, therefore allowing high accuracy to be achieved. On the other hand, the scattering term on the right hand side of Eq. (1) can be linearized when the overall temperature difference across the entire simulation domain is small relative to the absolute magnitude of the temperature. An important consequence of linearizing the scattering term is that the cumulative distribution function used to reset the states of scattered phonons does not depend upon the local equivalent equilibrium temperature, therefore eliminating the need to compute the local equivalent equilibrium temperature that is required in the conventional direct simulation MC scheme and also allowing the computational particles to be simulated sequentially[25]. Note that the form of the phonon BTE in Eq. (1) is presented in terms of the deviational energy distribution that allows the energy conservation to be accomplished naturally in the simulation.

For our application in this study to calculate the nanowire thermal conductivity, we specify the hot side and cold side temperatures of the simulated nanowires to be 301K and 300K, respectively. The reference temperature is chosen to be the same as the cold side temperature. As a result, phonons are only emitted from the hot boundary since the deviational energy on the cold side is exactly zero. The number of simulated phonon particles is determined by balancing the simulation accuracy with the computational cost. For all our simulations of nanowires of variable diameters, we use five million particles or more to achieve high accuracy with an acceptable computational cost. Since we solve the linearized deviational energy-based BTE, computational particles are simulated sequentially. The detailed algorithm to carry out the variance reduced MC simulation based on the deviational BTE is discussed extensively elsewhere[24,25,37] and we only briefly describe the procedure here. For each computational phonon, we first initialize its frequency, branch, and group velocity according to the cumulative distribution function determined by the hot side boundary temperature, as explained in ref. 23. The initial position of each computational particle is drawn randomly across the cross-sectional area at the hot boundary of the nanowire. The initial traveling direction is randomly generated with the direction pointing into the nanowire simulation domain to model the phonon emission process. We subsequently keep track of the trajectory of each individual particle and process anharmonic scattering and boundary scattering as needed until the particle gets absorbed by either the hot boundary or the cold boundary. If the particle ends up being absorbed by the cold boundary, it contributes to the heat flux; otherwise, it does not. The thermal conductivity of nanowires of variable diameters is determined by the heat flux contribution from all the phonon particles and the geometric parameters of the nanowires.

3. Simulation Results and Discussion

We apply the developed simulation framework to study the effect of phonon-boundary scattering on the thermal conductivity of square and circular PbSe nanowires under different boundary specularities[38]. The input phonon dispersion and lifetimes were calculated from first-principles density functional theory[26]. In this study, we are only interested in the classical size effect caused by the finite nanowire diameter rather than the nanowire length and thus we set the nanowire length to be much longer than the phonon MFPs that contribute to the thermal conductivity of PbSe, eliminating any non-diffusive transport effect caused by the finite nanowire length[39].

Figure 1 shows some representative temperature profiles along the axial direction of rough circular nanowires with three different diameters that span across the non-diffusive regime to diffusive regime. Rough nanowires refer to nanowires with diffuse boundary scattering, i.e. the boundary specularity $p = 0$. The temperature is averaged over the cross-sectional area at some discrete points along the nanowire. It is clear that the temperature distributions are all linear and no temperature jump is observed at the two ends of the nanowires, indicating that the nanowire length is sufficiently long to avoid the possible non-diffusive effect caused by the finite wire length[8]. We note that if the nanowire length were comparable with the phonon MFPs, the computed temperature distributions would have abrupt jumps at the two ends of the nanowires, as explained by prior numerical studies[8]. The linear temperature distribution in the non-diffusive transport regime also suggests that the nanowire has a well-defined effective diffusive thermal conductivity that may be much smaller than the bulk thermal conductivity value, as discussed in ref. 37 and in this work below.

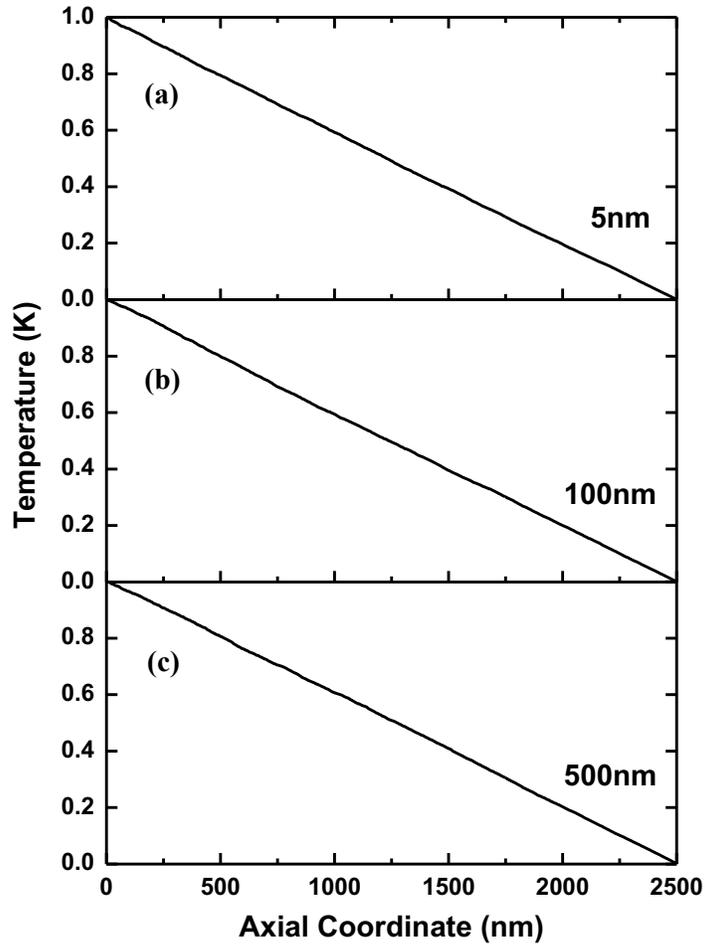

Figure 1. Axial temperature distribution for nanowires of three different diameters: (a) 5 nm, (b) 100 nm, and (c) 500 nm.

The diameter dependent thermal conductivities for rough PbSe nanowires are shown in Fig. 2. Results for both square and circular nanowires are included. As shown in Fig. 2, when the wire diameter is above 1 um, the computed effective thermal conductivity is close to the bulk value, suggesting negligible boundary scattering effect. However, as we systematically reduce the wire diameter below 1 um, we observe significant reduction in the lattice thermal conductivity, implying increasingly stronger boundary scattering effect with decreasing nanowire diameter. The suppression of thermal conductivity reaches ~ 40% for nanowire of about 10 nm in diameter. Such high reduction in thermal conductivity is helpful for improving thermoelectric energy conversion efficiency using PbSe nanostructures[2,3]. The simulation results also show the the wire shape has a very weak effect on the nanowire thermal conductivity. Instead, it is the wire size that determines the nanowire's heat transfer ability.

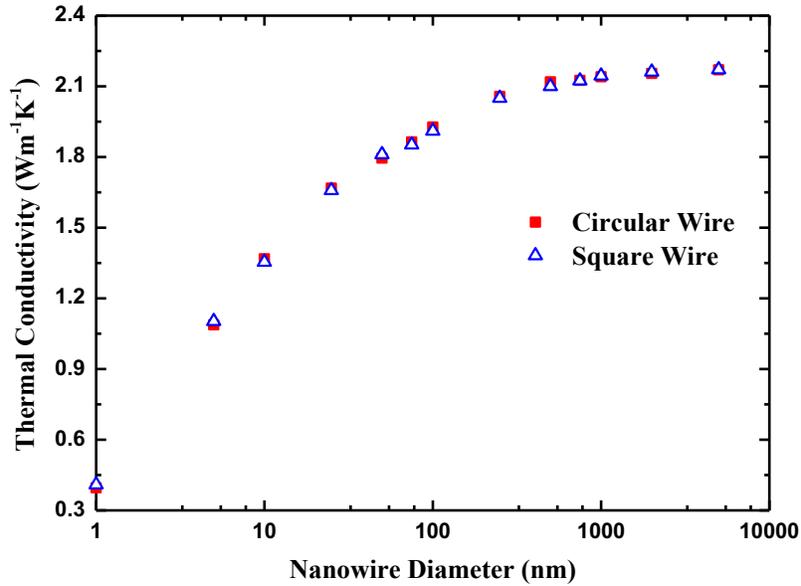

Figure 2. Computed PbSe nanowire thermal conductivities as a function of diameter for both square and circularly shaped nanowires. No significant shaped dependence is observed.

The role of the boundary specularity on the PbSe nanowire thermal conductivity is also systematically examined. Figure 3 shows the diameter-dependent thermal conductivities for nanowires with completely diffuse-scattering boundaries and partially diffuse and partially specular boundaries at room temperature (300K). The boundary specularity $p$ denotes the fraction of phonons that are specularly scattered at the wire boundary[8]. Since specular phonon reflection at the boundary does not cause additional thermal resistance for heat flow along the nanowire, we expect to observe higher thermal conductivities with more specular nanowires. Indeed, as shown in Fig. 3, the size effect is significantly weakened with increasing boundary specularity, consistent with our understanding of specular phonon-boundary scattering. When the boundary specularity parameter is unity (i.e., perfect specular reflection), our simulation correctly returns the bulk thermal conductivity for nanowire of any diameter (not shown here), again verifying the accuracy of the simulation results.

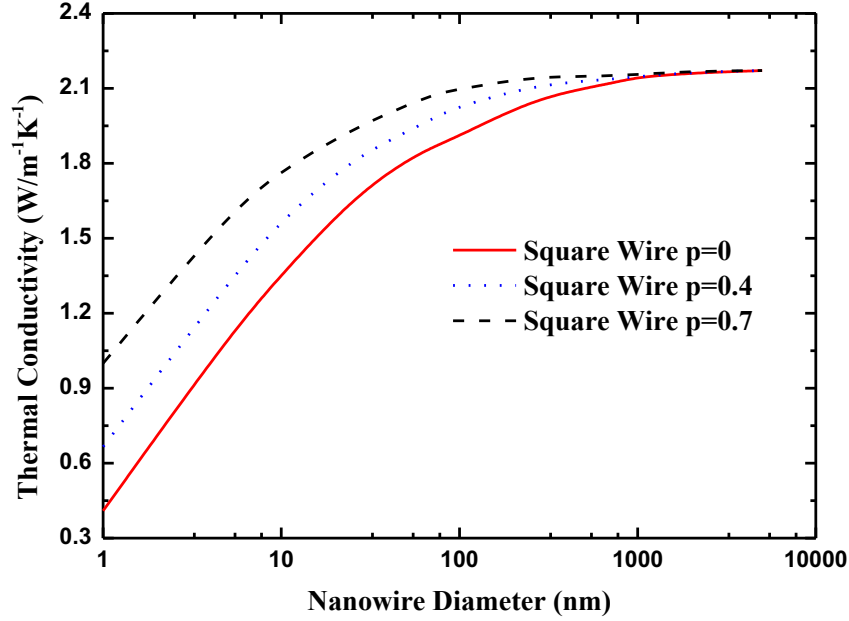

Figure 3. Computed size-dependent PbSe nanowire thermal conductivities across a range of boundary specularities.

4. Phonon MFP Reconstruction

Recent progress on the phonon MFP spectroscopy has shown that size-dependent thermal conductivities contain important information on the phonon MFP distributions that contribute to thermal transport in the material being studied[40–45]. Here, we follow the MFP reconstruction algorithm proposed by Minnich[46] to approximately extract the phonon MFP spectrum in PbSe using the computed diameter-dependent thermal conductivities. In general, the size-dependent thermal conductivities are connected to the phonon MFP distribution function through the following integral equation[46]:

$$k(\text{d}) = \int_0^\infty f(\Lambda) S(\eta) d\Lambda = -\int_0^\infty \frac{dS}{d\eta} \frac{d\eta}{d\Lambda} F(\Lambda) d\Lambda \qquad (2)$$

where $k(d)$ is the size-dependent thermal conductivity at the corresponding characteristic thermal length d (nanowire diameter for this study), $f(\Lambda)$ and $F(\Lambda)$ are the differential and cumulative MFP distribution functions, respectively, $S(\eta)$ is the suppression function, $\eta = \frac{\Lambda}{d}$ is the ratio of MFP to the characteristic thermal length d. The cumulative MFP distribution function $F(\Lambda)$ is a measure of the integral thermal conductivity contribution from phonons with MFPs less than $\Lambda$: $F(\Lambda) = \int_0^\Lambda f(\Lambda')d\Lambda'$. The suppression function is a measure of the magnitude of the size effect for a given sample configuration. The basic idea behind phonon MFP reconstruction is combine both the size-dependent thermal conductivities and the suppression function to inverse-solve the thermal conductivity accumulation function $F(\Lambda)$. In principle, this is mathematically doable. However, in practice, solving the integral Eq. (2) is generally ill-posed, as explained in detail in ref. 45. Fortunately, we can cast solving the integral Eq. (2) into a convex optimization problem by minimizing the following penalty function[46]:

$$P = ||AF - k||_2 + \lambda ||\Delta^2 F||_2 \qquad (3)$$

where A is a matrix related to the suppression function, $F$ is a vector of the desired thermal conductivity accumulation function at a set of discrete MFP values, $k$ is a vector of the size-dependent thermal conductivities, $\lambda$ is a smoothing parameter used to avoid abrupt jumps in the reconstructed MFP distribution function, $\Delta^2 F = F_{n+1} - 2F_n + F_{n-1}$, $||\ ||$ denotes the second norm. This casting process can naturally incorporate additional known constraints on $F(\Lambda)$. For example, we can include the fact that the thermal conductivity accumulation increases monotonically from zero to unity by adding the constraints $F_1 = 0, F_{n+1} \geq F_n (n = 1,2,\ldots, N-1), F_N = 1$, where $N$ is the total number of discrete MFP values across which the thermal conductivity accumulation function varies from zero to unity.

In practice, since most microfabricated nanowires have diffuse boundaries, we use the thermal conductivities of the rough nanowires to perform the MFP reconstruction. The corresponding suppression function was derived by Dingle[29] and is plotted in Fig. 4. This suppression function describes the reduction of nanowire thermal conductivity due to phonon-boundary scattering in the non-diffusive regime and is similar in essence to the Fuchs-Sondheimer[27,28] type of suppression function for the thermal conductivity of thin membranes along the in-plane direction.

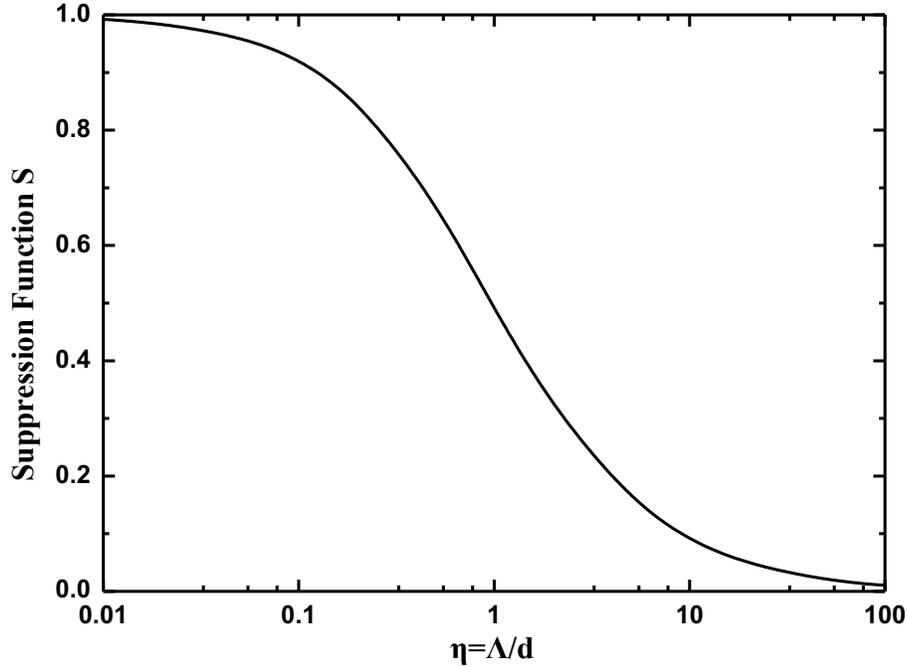

Figure 4. Thermal conductivity suppression function of nanowire as a function of the ratio of MFP to the nanowire diameter d. ≈η=Λ/d

As pointed out by ref. 43, the smoothing parameter $\lambda$ should have the minimum value that still does not give abrupt jumps in the reconstructed thermal conductivity accumulation function. In this study, we find that an appropriate value for the smoothing parameter is ~ 0.5. Figure 5 shows the comparison between the reconstructed MFP distribution and the input MFP distribution that was used to compute the diameter-dependent thermal conductivities. As shown in Fig. 5, the reconstructed phonon MFP distribution agrees excellently with the input thermal conductivity accumulation function. This agreement suggests phonon MFP distributions of various materials could, in principle, be determined from thermal conductivity measurement of nanowires across an array of diameters.

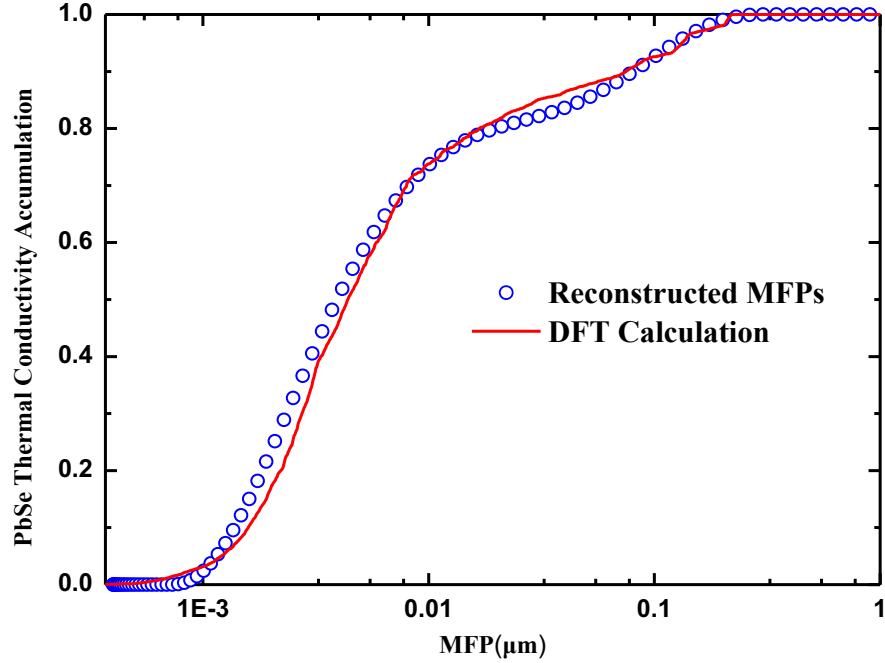

Figure 5. Comparison of reconstructed PbSe phonon MFP distributions with DFT calculated result.

5. Conclusion

In summary, we rigorously solve the multidimensional spectral phonon Boltzmann transport equation to examine the impact of phonon-nanowire boundary scattering and boundary specularity on the thermal conductivity of PbSe nanowires. For rough nanowires, significant reduction in thermal conductivity is observed for PbSe nanowires with diameters smaller than a few hundred nanometers, implying potential ability to further improve the thermoelectric energy conversion efficiency through nanostructuring PbSe. Increasing boundary specularity results in increase in thermal conductivity due to less thermal resistance caused by the boundary scattering. We also find, through algorithmic reconstruction, that the diameter-dependent nanowire thermal conductivity contains important information about the phonon mean free path distribution in the material under study. Our simulation results will help gain fundamental insight into nanoscale thermal transport in PbSe nanowires and also assist in size-effect engineering in thermoelectric applications.

Author Contributions

L.M. and R.M contributed equally to this work.

Financial Interests

The authors declare no competing financial interests.

Acknowledgement

This work is supported by National Natural Science Foundation of China (No. 51306125) and by Shenzhen Research Foundation of Science & Technology (KQCX20140519105122378, JCYJ20130329113322731).